# Discrete Element Parameter Calibration of Livestock Salt Based on Particle Scaling

Lulu Nie[a,b], Baoqin Wen[a,b,*], Jingbin Li[a,b], Shufeng Li[a,b], Yali Li[a,b], Zhaokun Zhang[a,b], Zhiyuan Wang[a,b], Zhihao Fan[a,b]

[a] *College of Mechanical and Electrical Engineering, Shihezi University, Shihezi 832003, China*

[b] *Xinjiang Production and Construction Corps Key Laboratory of Modern Agricultural Machinery, Shihezi 832003, China*

**TheoryAbstract:** In order to obtain accurate contact parameters for the discrete element simulation of salt particles used in animal husbandry, the principle of particle contact scaling and dimensional analysis were used for particle scaling. Firstly, the Plackett Burman experiment was used to screen the parameters that significantly affect the angle of repose: salt salt rolling friction coefficient, salt salt recovery coefficient, and salt steel rolling friction coefficient. Considering the influence of other parameters, a combination of bench and simulation experiments was used to calibrate the contact parameters between salt particles and steel plates used in animal husbandry in EDEM. Finally, through the stacking test, steepest climbing test, and orthogonal rotation combination test, the salt salt rolling friction coefficient was obtained to be 0.23, the salt salt recovery coefficient was 0.544, and the salt steel rolling friction coefficient was 0.368, which were verified through bench tests. The experimental results show that the relative error between the actual value of the stacking angle and the simulation results is 0.6%. The results indicate that the calibrated contact parameters can be used for discrete element simulation of salt particles for animal husbandry, providing reference for the design of quantitative feeding screws and silos.

**0 foreword**

Mineral elements are essential components of animal tissues and organs, with salt (sodium chloride) serving as a core mineral constituent. Its content directly impacts livestock health, as insufficient sodium and chlorine in feed can reduce appetite in sheep, and prolonged deficiencies may lead to lethargy, diminished vitality, and impaired growth[1]. Modern livestock farming employs screw conveyor systems for automated salt delivery and metering, where the scientific design of these systems critically determines salt transport efficiency and dosing precision.

The primary challenge in simulation experiments lies in the precise calibration of parameters, where the accuracy of parameters for livestock salt particles is critical to the reliability of simulation results. Currently, the Discrete Element Method (DEM) has been widely applied to model fine particulate materials (e.g., ores, feed, and soil), with extensive

E–mail addresses: 20232109096@stu.shzu.edu.cn(Lu N)
*Corresponding author.
E–mail addresses:wbq1980@shzu.edu.cn (Wen B).

research conducted by scholars globally. Significant advancements in DEM parameter calibration include:Zhang et al[2] calibrated scaled-up particle parameters using DEM to match the flow characteristics of real fine iron tailings, enabling accurate simulations.Guo Sanqin et al.[3] combined simulations and bench tests to calibrate contact parameters between pelletized feed and steel/nylon surfaces, supporting precision feeding technologies.Ucgul et al. [4] determined contact models and parameters for non-cohesive soils through angle of repose and penetration tests, effectively simulating soil tillage processes.Yang She et al. [5] established a DEM model for sand particles by comparing physical and simulated experiments, laying a foundation for mechanical sand removal studies.Soltanbeigi et al[6] simulated the stacking of spherical glass particles, identifying sliding and rolling friction coefficients between particles and container walls as key factors influencing pile morphology.Han Wei et al. [7] employed the JKR contact model to calibrate parameters for micron-scale reactive dye particles, offering insights for refined simulations.These studies provide critical guidance for DEM applications in particulate material simulations, particularly in agriculture, mining, and chemical engineering. Sun et al. [8] derived curve equations and throughput formulas for screw feeders under varying friction coefficients through DEM simulations, enhancing feeder design efficiency. Ren Jianli et al. [9] validated particle scaling theory by simulating cast iron-coal particle motion in vertical screw conveyors. Weinhart et al. [10] demonstrated the importance of spatial coarse-graining and temporal intervals in DEM simulations of silos using coarse-grained models.

As a specialized solid granular material, livestock salt particles require systematic acquisition of their physical characteristic parameters through DEM-based simulation modeling and parameter calibration. This process provides a theoretical basis for the optimized design of screw conveyor systems and the adjustment of operational parameters. Accordingly, this study focuses on livestock salt particles, establishing a spherical particle DEM model using EDEM software. The research objectives include:

(1) Calibrating contact parameters for livestock salt particles by comparing bench tests with simulation experiments;

(2) Determining interparticle contact parameters through stacking tests, steepest ascent experiments, and quadratic regression orthogonal rotational composite experiments, optimized via response surface methodology (RSM).

This work aims to deliver critical data support for the precision design and efficient operation of livestock salt conveying systems.

# 1 Materials and Methods
## 1.1 Acquisition of Experimental Materials

The experimental material consisted of livestock salt produced by Golmu Salt Chemical Co., Ltd., compliant with the Chinese national standard GB/T21513. Since salt primarily interacts with metal surfaces during conveying processes, steel was selected as the contact material. The physical parameters of steel were determined as follows: Poisson's ratio = 0.25, density = 1210 kg/m³ , and shear modulus = 1900 MPa.

## 1.2 Particle Scaling Principles

## 1.2.1 Dimensional Analysis

To improve the accuracy and effectiveness of discrete element method (DEM) simulations, optimization of simulation parameters is required to ensure that scaled-up particle models faithfully replicate the dynamic and static behavioral characteristics of real particle systems, thereby effectively mitigating simulation deviations caused by particle size scaling. The generalized governing equations of particle motion proposed by Feng et al. [11] can be expressed as:

$$m\ddot{u}(t) + F_d(t) + F_{int}(t) = F_{ext}(t) \quad (1)$$

where: m is the particle mass; $F_d$ represents the damping force accounting for energy dissipation within the system; $F_{int}$ denotes the resultant interaction force from other particles or phases defined by contact laws; $F_{ext}$ is the externally applied resultant force; u is the acceleration, $\ddot{u}$ can be interpreted as the derivative of displacement/position.。

A relatively simple method was proposed by Feng et al. [11] to establish a set of scaling laws by ensuring proportional equivalence of all corresponding forces between the two models.

$$\frac{\bar{m}\ddot{\bar{u}}}{m\ddot{u}} = \frac{\bar{F}_d}{F_d} = \frac{\bar{F}_{int}}{F_{int}} = \frac{\bar{F}_{ext}}{F_{ext}} = \lambda \quad (2)$$

$$\bar{q} = \lambda_q * q \quad (3)$$

where: q denotes an arbitrary parameter in the physical system; $\lambda_q$ is the scaling factor; q represents the corresponding parameter in the scaled system.

When all scaling factors for physical quantities are determined, a scaled model can be established. In the original system, except for a small number of independent fundamental quantities, all other physical quantities can be derived from these fundamental quantities. While mass density [ρ] replaces mass (m) as a fundamental quantity, length (L) and time (T) remain as the other two fundamental quantities. The scaling factors for these three quantities $\lambda_L$、$\lambda_M$、$\lambda_\rho$ are specifically selected as follows:

During the establishment of a discrete element method (DEM) scaling model, a complete scaled system can be constructed once the scaling factors for all relevant physical quantities are determined. In this system, apart from the three independent fundamental quantities—length (L), mass (M), and mass density (ρ)all other physical quantities can be derived through dimensional relationships among these fundamental quantities. The scaling factors for these three fundamental quantities are defined as follows:

$$\lambda_L = h;\ \lambda_M = h;\ \lambda_\rho = 1 \quad (4)$$

The optimal method for calculating the scaling factor of a quantity involves expressing its dimensional formula in terms of fundamental quantities and then converting these dimensions into scaling factors. For example, the mass M can be expressed as M=ρV, where ρ is the mass density and V is the volume.

$$[M] = [\rho][L]^3; \lambda_M = \lambda_\rho \lambda_L^3 = h^3 \quad (5)$$

For force (F), the scaling factor is derived using Newton's second law ($F=ma$)

$$[F] = [M][L][T]^{-2} = [\rho][L]^4[T]^{-2}; \lambda_F = \lambda_\rho \lambda_L^4 \lambda_T^{-2} = h^2 \quad (6)$$

For Young's modulus (E), consider the relationship between axial force (F) and axial displacement (u) for a one-dimensional rod with Young's modulus E, cross-sectional area A, and length L:

$$F = \frac{EA}{L} u \quad (7)$$

This yields:

$$[E] = [\rho][L]^2[T]^{-2}; \lambda_E = \lambda_\rho \lambda_L^2 \lambda_T^{-2} = 1 \quad (8)$$

The dimensional analysis results demonstrate that the scaled model system exhibits the following key characteristics: density remains invariant throughout the scaling process; contact stiffness displays size dependency, scaling linearly with particle diameter; and elastic modulus serves as a dynamic parameter, adaptively calibrated in response to particle size variations. For parameter selection, the upper bounds of parameter intervals are prioritized to ensure numerical stability.

1.2.2 Scaling AnalysisStefan

Radl S et al. [12] analyzed the linear elastic model by ensuring identical rotational kinetic energy between the original and scaled systems. According to Newton's laws of motion, the differential equation governing the normal overlap between particles can be expressed as:

$$m_e \delta_n'' = k_n \delta_n + c_n \delta_n' \quad (9)$$

In the formula $m_e$ is the effective mass,kg;$k_n$ for particle stiffness,N/m;$c_n$ for damping coefficient,kg/s;$\delta_n$ for overlap quantity,m。

$$m_e = \frac{4\pi R_i^3 \rho \beta^3}{3(1+\beta^3)} \quad (10)$$

where: $\rho$ is the particle density (kg/m³);$\beta$ is the particle size ratio;$R_i$ is the particle radius in the original system.

$$R_e = \frac{R_i R_j}{(R_i + R_j)} = \frac{R_i \beta}{1+\beta} \quad (11)$$

In the formula $R_e$ is the effective radius,m;$R_j$ is the particle radius of the scaling system,m;convert equations (10) and (11) 2 into dimensionless quantities。

$$\delta_n^* = \delta_n/R; \delta_n'^* = \delta_n'/v_0; t^* = t/(R_i/v_0) \Rightarrow \delta_n''^* = \delta_n''/(R_i/v_0^2) \quad (12)$$

In the formula * is a dimensionless number,$\delta_n$ is the overlap,m;t is time,s;$v_0$ is velocity,m/s。

By substituting into equation (10), we can obtain

$$\frac{4\pi R_i^2 \rho \beta^3 v_0^2}{3(1+\beta^3)} \delta_n''^* = k_n R_i \delta_n^* + c_n v_0 \delta_n'^* \tag{13}$$

In the formula, $k_n$ represents the particle stiffness, N/m; $c_n$ is the damping coefficient, kg/s。Simplify to obtain

$$\frac{4\pi \beta^3}{3(1+\beta^3)} \delta_n''^* = \frac{k_n \delta_n^*}{R_i \rho v_0^2} + \frac{c_n \delta_n'^*}{R_i^2 \rho v_0} \tag{14}$$

The coefficients in equation (14) can be represented by the following dimensionless numbers:

$$\pi_1 = \frac{4\pi \beta^3}{3(1+\beta^3)}; \pi_2 = \frac{k_n}{R_i \rho v_0^2}; \pi_3 = \frac{c_n}{R_i^2 \rho v_0} \tag{15}$$

When scaling the system, invariance in density and velocity must be maintained. In the scaled system, the relative overlap between particles remains constant, denoted as the dimensionless parameter $\pi_1 = \lambda_1$ which necessitates preserving the particle size ratio. The invariance $\pi_2 = \lambda_2$ and $k_n/R_i^2$ confirms the linear dependence of stiffness on radius, while $\pi_3 = \lambda_3$ and $c_n/R_i^2$ establish that the coefficient of restitution is proportional to the square of the radius. For low-cohesion materials, variations in flow regimes caused by increased solid fraction are neglected in this study, aligning with the assumptions of Model [13]. The coefficient of restitution, a contact-dependent parameter, must be calibrated to account for scaling effects. To minimize simulation errors and optimize computational efficiency, the livestock salt particles were scaled by a factor of 2 for simulations, consistent with methodologies in prior studies [14].

## 2 Parameter Calibration Process

### 2.1 Physical Parameters

#### 2.1.1 Angle of Repose Measurement

This study followed the GB/T standard method and incorporated existing literature on the angle of repose[15, 16, 17]. The angle of repose of livestock salt was measured using the funnel method. In accordance with the GB/T 16913.5-1997 standard, the angle of repose was determined using a BT-1000 powder characteristic tester and the injection method (funnel method). The salt particles were slowly poured through a funnel, with a glass rod used to prevent clogging. After the pile stabilized, the angle was measured using the tester's repose angle module (Fig.1). The experiment included five replicates, each with five parallel measurements. The average angle of repose was determined to be 45.27° (see Table 1 for details).

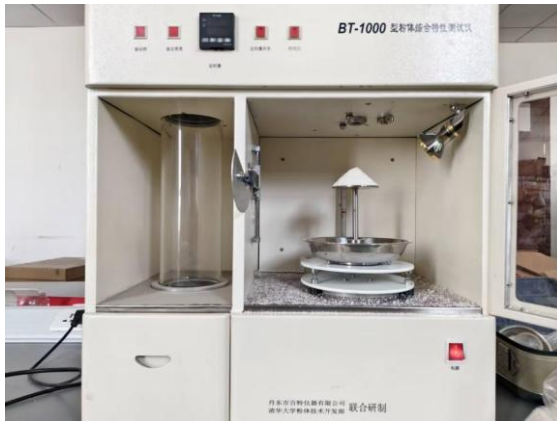
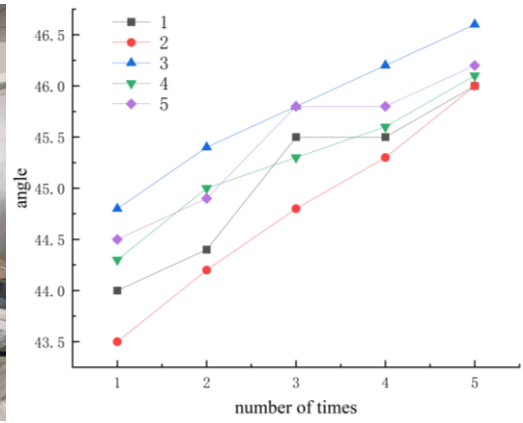

Figure 1 BT-1000 Powder Comprehensive Characteristics Tester

Table 1 Particle angle of repose

| group count | angle | | | | | mean value |
|---|---|---|---|---|---|---|
| 1 | 44 | 44.4 | 45.5 | 45.5 | 46 | 45.08 |
| 2 | 43.5 | 44.2 | 44.8 | 45.3 | 46 | 44.76 |
| 3 | 44.8 | 45.4 | 45.8 | 46.2 | 46.6 | 45.76 |
| 4 | 44.3 | 45 | 45.3 | 45.6 | 46.1 | 45.26 |
| 5 | 44.5 | 44.9 | 45.8 | 45.8 | 46.2 | 45.44 |
| mean value | | | | | | 45.27 |

2.1.2 Geometric Size Distribution

To determine the precise size range of salt particles, a five-point sampling method was employed, randomly collecting a 500 g sample from the purchased bag of salt. The sample was sieved using mesh screens of varying apertures [18] (Fig. 2), and the resulting particle size distribution is presented in Table 2. The size distribution range of salt particles was calculated using the following formula: [Insert Formula].

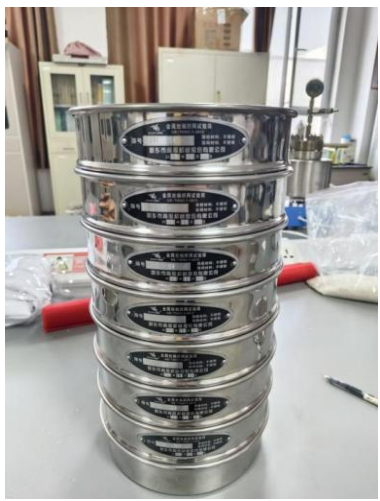
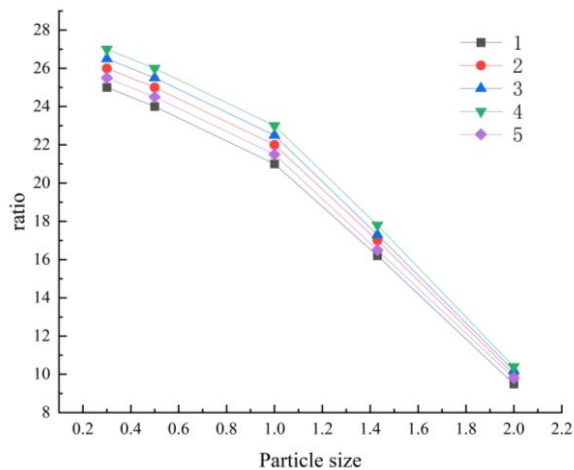

Figure 2 Particle size experiment

Table 2 Grain size of salt for animal husbandry

| group count | 0.3mm | 0.5mm | 1mm | 1.43mm | 2mm |
|---|---|---|---|---|---|
| 1 | 25.0% | 24.0% | 21.0% | 16.2% | 9.6% |
| 2 | 25.5% | 24.5% | 21.5% | 16.5% | 9.8% |
| 3 | 26.0% | 25.0% | 22.0% | 17.0% | 10.0% |
| 4 | 26.5% | 25.5% | 22.5% | 17.3% | 10.2% |
| 5 | 27.0% | 26.0% | 23.0% | 17.8% | 10.4% |
| mean value | 26% | 25% | 22% | 17% | 10% |

2.1.3 Density Measurement

This experiment determined the density of salt particles using the volume displacement method: First, saturated brine was prepared by continuously adding salt to 50 mL of distilled water until dissolution equilibrium was reached. The total mass of the salt-beaker-brine system was measured using an electronic balance. A measured quantity of saturated brine was then poured into a graduated cylinder, and the initial volume ($V_1$) was recorded. After adding salt particles and removing air bubbles, the final volume ($V_2$) was recorded. Using the density formula (Formula 16), the density of salt was calculated to be 1210 kg/m³ (see Table 3 for details).

$$\rho = \frac{m}{V_2 - V_1} \quad (16)$$

Table 3 Salt Density for Livestock Use

| number | M/g | V/cm³ | ρ(kg/m³) |
|---|---|---|---|
| 1 | 400 | 328 | 1219 |
| 2 | 400 | 327 | 1223 |
| 3 | 400 | 330 | 1212 |
| 4 | 400 | 333 | 1201 |
| 5 | 400 | 332 | 1204 |
| mean value | | | 1210 |

**2.2 Simulation Model**

2.2.1 Simulation Parameters

Based on literature [**19, 20, 21, 22, 23, 24**] and the GEMM material database in EDEM software, the range of particle-stainless steel contact parameters was determined (Table 5). The intrinsic density of salt particles was set as 1210 kg/m³ according to particle scaling theory (data from Table 4). Since contact parameters (friction/restitution coefficients) are affected by both material properties and geometric morphology and cannot be directly obtained from physical properties, they were determined through virtual experiments using inverse parameter calibration. The Hertz-Mindlin no-slip contact model [**25**] was selected for EDEM simulations. This model, based on Mindlin's theoretical work, offers accurate and efficient computational performance and effectively simulates contact behavior between granular materials.

To address the lack of material property parameters for salt particles: while intrinsic material parameters closely match real values, direct measurement of contact parameters through bench tests may deviate from actual values due to the small particle size, necessitating contact parameter recalibration [**26**]. This study developed a parameter calibration method based on angle of repose tests: screening key contact parameters through significance analysis

[27];calibrating secondary parameters with bench tests [28]; and optimizing interparticle contact parameters using relative angle of repose error as the indicator, through steepest ascent experiments for main effect parameters and quadratic regression orthogonal rotation combination design.

2.2.2 Simulation Model

Following GB/T 11986-98 standards, a discrete element angle of repose simulation model was established (Fig. 3), replicating experimental conditions with injection method: funnel orifice diameter 20 mm, base plate diameter 80 mm, and discharge height 75 mm.Particle parameters included spherical particles [29, 30, 31, 32, 33], generation rate of 10,000 particles/s, total 10,000 particles, and simulation duration 12 s, with angle of repose directly measured using the software protractor module.

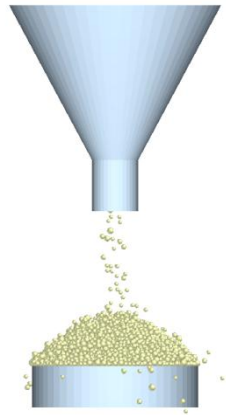

Figure 3 Simulation Experiment

Table 4: Intrinsic parameters of salt and geometry

| Simulation parameters | numerical value |
| --- | --- |
| Salt density/(kg • m²) | 1210 |
| Poisson's ratio of stainless steel | 0.30 |
| Density of stainless steel/(kg • m²) | 7800 |
| Shear modulus/Pa of stainless steel/Pa | $8.0 \times 10^{10}$ |

Table 5: Contact Parameters of Salt and Geometric Bodies

| Factor number | factor name | low level | high tone |
| --- | --- | --- | --- |
| A | Salt - steel recovery coefficient | 0.3 | 0.75 |
| B | Salt - steel static friction system | 0.35 | 0.9 |
| C | Salt - steel rolling friction | 0.2 | 0.5 |
| D | Poisson's ratio of salt | 0.2 | 0.35 |
| E | Salt - salt recovery coefficient | 0.15 | 0.75 |
| F | Salt - salt static friction system | 0.4 | 0.9 |
| G | Salt - salt rolling friction | 0.05 | 0.35 |

**2.3 Response Surface Methodology for Simulation Parameters**

2.3.1 Plackett-Burman Design

The Plackett-Burman design was employed to screen significant contact parameters, with the angle of repose as the response variable. Parameter levels were set with the low level as baseline values and the high level as twice the baseline. Key significant parameters were identified by analyzing angle of repose differences between these levels.Analysis of variance (ANOVA) results from Design Expert for the Plackett-Burman design (Table 6) demonstrated that the salt-steel rolling friction coefficient, salt-salt restitution coefficient, and rolling friction coefficient (P < 0.01) had highly significant effects on the angle of repose, while other parameters (P > 0.05) showed no significant influence.Consequently, only these three significant parameters were optimized in subsequent steepest ascent and Box-Behnken designs, while non-significant parameters were determined through a combined bench test-simulation calibration approach.

Table 6 Plackett Burman experimental design and results

|    | A  | B  | C  | D  | E    | F  | G  | angle of repose |
|----|----|----|----|----|------|----|----|-----------------|
| 1  | +1 | +1 | -1 | +1 | 0.75 | +1 | -1 | 25.91           |
| 2  | -1 | +1 | +1 | -1 | 0.75 | +1 | +1 | 46.23           |
| 3  | +1 | -1 | +1 | +1 | -1   | +1 | +1 | 55.12           |
| 4  | -1 | +1 | -1 | +1 | 0.75 | -1 | +1 | 40.12           |
| 5  | -1 | -1 | +1 | -1 | 0.75 | +1 | -1 | 31.25           |
| 6  | -1 | -1 | -1 | +1 | -1   | +1 | +1 | 46.85           |
| 7  | +1 | -1 | -1 | -1 | 0.75 | -1 | +1 | 43.26           |
| 8  | +1 | +1 | -1 | -1 | -1   | +1 | -1 | 34.56           |
| 9  | +1 | +1 | +1 | -1 | -1   | -1 | +1 | 51.93           |
| 10 | -1 | +1 | +1 | +1 | -1   | -1 | -1 | 35.62           |
| 11 | +1 | -1 | +1 | +1 | 0.75 | -1 | -1 | 31.24           |
| 12 | -1 | -1 | -1 | -1 | -1   | -1 | -1 | 31.58           |

Table 7: Significance Analysis of Plackett Burman Test Parameters

|                              | The mean square sum |   | Mean Square | f-value | P value  |             |
|------------------------------|---------------------|---|-------------|---------|----------|-------------|
| Model                        | 930.39              | 7 | 132.91      | 104.39  | 0.0002   | significant |
| A Salt-Steel Recovery        | 8.96                | 1 | 8.96        | 7.04    | 0.0568   |             |
| B Salt-Steel Static Friction | 2.03                | 1 | 2.03        | 1.59    | 0.2758   |             |
| C salt-steel rolling friction| 70.62               | 1 | 70.62       | 55.46   | 0.0017   |             |
| D Salt Poisson's ratio       | 1.30                | 1 | 1.30        | 1.02    | 0.3694   |             |
| E Salt - Salt Recovery       | 118.13              | 1 | 118.13      | 92.78   | 0.0006   |             |
| F Salt-Salt Static Friction  | 3.17                | 1 | 3.17        | 2.49    | 0.1896   |             |
| G Salt-Salt Rolling Friction | 726.19              | 1 | 726.19      | 570.37  | <0.0001  |             |

G-Salt Salt Rolling Friction

The Plackett-Burman design identified three significant parameters: salt-salt rolling friction coefficient, salt-salt restitution coefficient, and salt-steel rolling friction coefficient. Although other parameters did not reach significance (p ≥ 0.05), they may still potentially influence the angle of repose. Therefore, their optimal values were determined through inverse parameter identification using response surface fitting with a combined bench test-simulation calibration approach.

2.3.2 Static Friction Coefficient Calibration

The static friction coefficient ($\mu$) was measured using an inclined plane sliding test. This method was employed to calibrate the static friction coefficient between livestock salt particles and stainless steel plate, with the experimental setup shown in Figure 4. The calculation formula is:

$$\mu = \frac{mg\sin\alpha}{mg\cos\alpha} = \tan\alpha \tag{20}$$

in the formula

α————Inclination angle of sliding plane，(°)

m————Mass of livestock salt particles（kg）

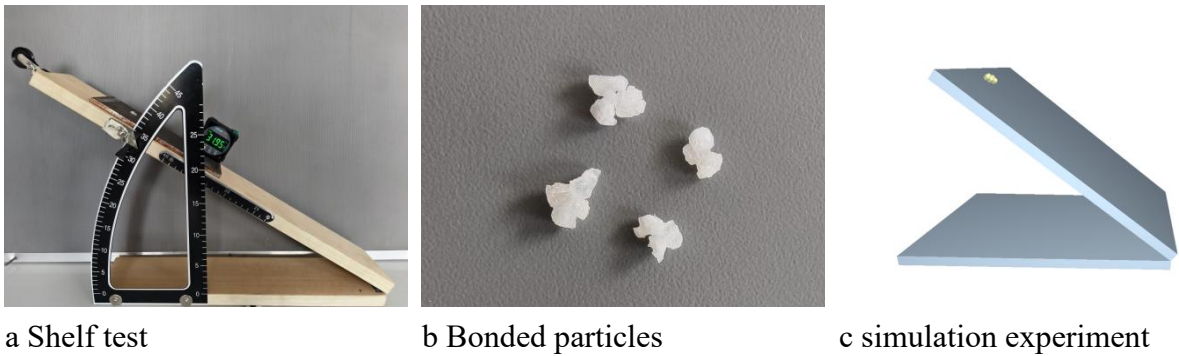

a Shelf test　　　　　　　b Bonded particles　　　　　c simulation experiment

Figure 4 Calibration experiment of static friction coefficient

In this experiment, four livestock salt particles (prevented from rolling) were bonded to a horizontal stainless steel plate. The plate was then tilted at a constant angular velocity of 5°/s, and the inclination angle was recorded when the particles began to slide. The average sliding angle was determined from five repeated trials.Since the salt-steel rolling friction coefficient and salt-salt contact parameters (restitution coefficient, static friction coefficient , and rolling friction coefficient ) had no significant effect on the sliding angle, these values were set to zero in the EDEM simulations.The simulation model was configured as follows: An 80 mm×60 mm double-layer stainless steel plate was imported, with particles generated 3 mm above the upper plate. After 1 s, the upper plate began rotating at 5°/s for a total duration of 10 s, with a time step of $2.3 \times 10^{-6}$ s and a mesh size three times the minimum particle radius.Six simulation groups were conducted for the salt-steel static friction coefficient (ranging from 0.35 to 0.85 in increments of 0.1), with each group comprising five parallel trials. The average inclination angles are presented in Table 8.A quadratic polynomial regression model (Equation 21, fitted curve in Figure 5) was established based on the experimental data in Table 8 to describe the relationship between the salt-steel static friction coefficient and the inclination angle. The coefficient of determination ($R^2$ = 0.9986, approaching 1) indicates that the model accurately characterizes the parametric relationship.

Table 8 Static Friction Coefficient Simulation Test Plan and Results

| Number | Static Friction Coefficien | Angle |
|---|---|---|
| 1 | 0.35 | 19.97 |
| 2 | 0.45 | 24.97 |
| 3 | 0.55 | 29.97 |
| 4 | 0.65 | 33.43 |
| 5 | 0.75 | 37.46 |
| 6 | 0.85 | 39.96 |

The curve equation shown in the figure is

$$y = -33.71x^2 + 79.99x - 3.99 \tag{21}$$

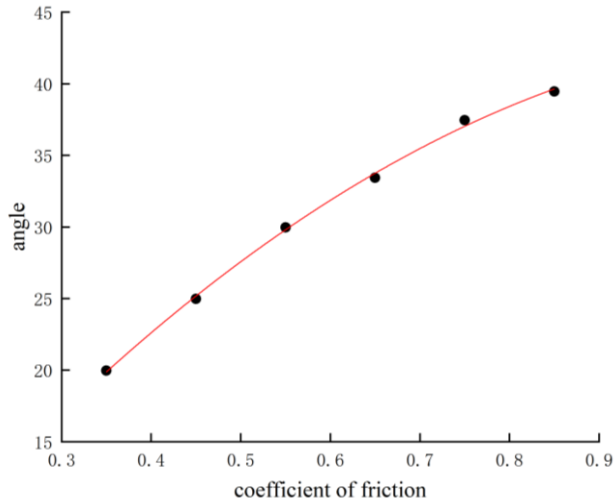

Figure 5 Fitting curve of static friction coefficient and tilt angle

By substituting the experimentally measured tilt angle of 35.82° into Equation (12), the salt-steel static friction coefficient was inversely calculated as 0.71. EDEM simulations (five repeated trials, mean angle = 36.01°) confirmed this value, showing a relative error of only 0.5% compared to the experimental measurement. Thus, the salt-steel static friction coefficient in EDEM was determined to be 0.71.

2.3.3 Calibration of the Restitution Coefficient

The restitution coefficient (e) is a physical parameter that characterizes energy transfer and rebound behavior during collisions. In this study, the restitution coefficient (e) between livestock salt particles and the stainless steel plate was calibrated using collision bounce tests. The restitution coefficient is calculated as:

$$e = \sqrt{\frac{h_1}{H_1}} \tag{22}$$

In the formula

$h_1$———Maximum rebound height of livestock salt particles (mm), mm

$H_1$———Initial release height of livestock salt particles, mm

In this experiment, salt particles were freely dropped from a height of 150 mm onto a horizontal stainless steel plate. The average maximum rebound height ($h_1$ = 20.5 mm) was obtained from five repeated trials. Since the salt-steel static/rolling friction coefficients and salt-salt contact parameters had negligible effects on rebound height, these values were set to zero in EDEM simulations.Simulation setup: Particles were dynamically generated 150 mm above a 100 mm × 60 mm stainless steel plate. The total duration was 2 s with a time step of $1.67\times10^{-5}$s, and the mesh size was three times the minimum particle radius.Seven simulation groups were conducted for the salt-steel restitution coefficient ($A_1$ = 0.3 - 0.75, increment 0.05), with each group comprising five trials. The average rebound heights ($B_1$) are summarized in Table 9.

Table 9 Simulation Test Scheme and Results of Collision Recovery Coefficient

| Number | recovery coefficient | height |
| --- | --- | --- |
| 1 | 0.35 | 16.2 |
| 2 | 0.40 | 20.16 |
| 3 | 0.45 | 22.55 |
| 4 | 0.5 | 30.26 |
| 5 | 0.55 | 42.44 |
| 6 | 0.6 | 46.59 |
| 7 | 0.65 | 58.75 |
| 8 | 0.7 | 68.92 |
| 9 | 0.75 | 87.35 |

To establish the relationship between the restitution coefficient and maximum rebound 9 for livestock salt particles impacting steel plates in simulations, the experimental data from Table 9 were fitted using a quadratic polynomial regression model. The resulting fitted curve demonstrates

$$y = 328.955x^2 - 186.957x + 40.9209 \qquad (23)$$

As shown in Figure 6, the curve equation is expressed as:

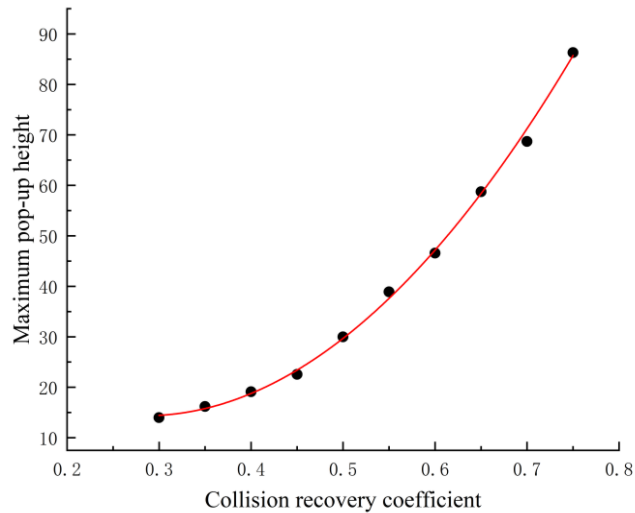

Figure 6 Curve equation of recovery coefficient

Equation (23) yields a coefficient of determination ($R^2$) of 0.998, indicating excellent model fitting accuracy.By substituting the experimentally measured rebound height of 20.5 mm into the equation, the restitution coefficient $A_1$ was determined to be 0.421 through inverse calculation, which was then verified via EDEM simulations (five replicate trials).The mean maximum rebound height from five simulation trials was 22.74 mm, showing only 1.09% deviation from bench-scale test results. Thus, the salt-steel restitution coefficient $A_1$ in EDEM was ultimately set as 0.421.

2.3.4 Determination of Salt-Salt Static Friction Coefficient

A ZJ-type direct shear apparatus was employed to measure the internal friction angle of granular materials. The yield locus was obtained through transient shear tests, from which the internal friction angle was derived. The testing protocol comprised three stages: pre-compaction, pre-shearing, and shearing. Let $m_a$ denote the combined mass of the material above the shear plane, shear ring, and shear cover after shearing. The normal stresses during pre-shearing ($W_p$) and shearing ($W_s$) stages were calculated as follows:

$$W_p = (m_h + m_p + m_a)g \qquad (24)$$

$$W_s = (m_h + m_s + m_a)g \qquad (25)$$

$W_p$———Positive pressure during the pre shearing stage,N;
$W_s$———Positive pressure during the shearing stage,N;
$m_h$———Quality of the hanger,kg;
$m_p$———Weight mass during pre cutting,kg;
$m_s$———Weight mass during cutting,kg;

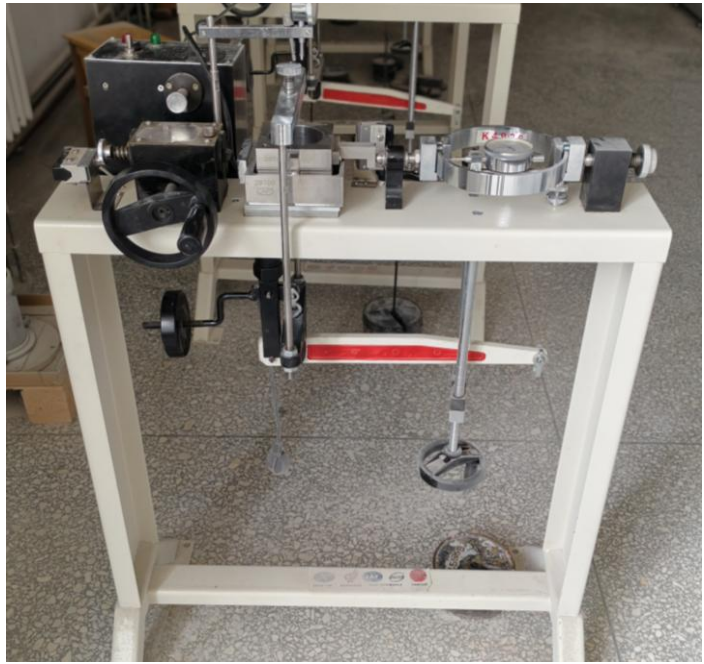

Figure 7 ZJ type direct shear tester

Under a constant pre-shear normal stress ($W_p$), transient shear tests were conducted by applying three or more different levels of shear normal stress ($W_s$), yielding the failure envelope shown in Figure 8. The measured shear forces ($F_{Si}$) and normal pressures ($W_{Si}$) were converted to shear stress $\sigma_s$ (kPa) and normal stress $\sigma_t$ (kPa). The internal friction angle was determined from the slope of the fitted curve in Equation (12), and subsequently transformed to the static friction coefficient through the conversion relationship in Equation (13), ultimately obtaining a salt-salt static friction coefficient of 0.85.

$$y=0.845x+0.084 \quad (26)$$

$$\mu_s = \tan\phi \quad (27)$$

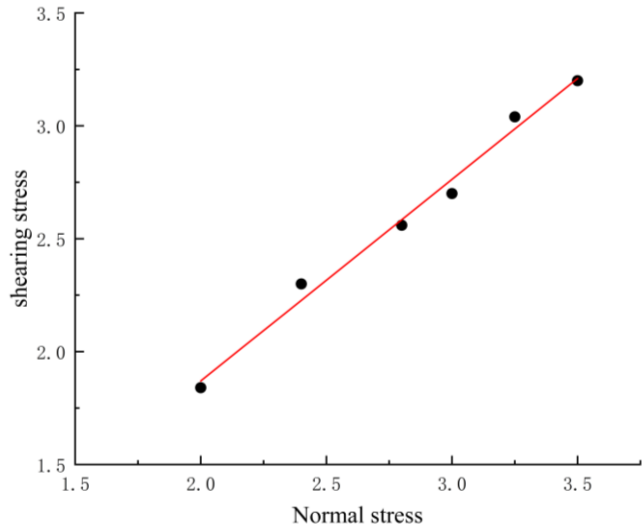

Figure 8 Yield trajectory of salt

2.3.5 Steepest Ascent Experiment Design and Results

Based on Plackett-Burman screening results, the steepest ascent method was employed to rapidly approach the optimal parameter region. Starting from the center point of PB experiments, larger step sizes were set according to regression coefficients to accelerate convergence (experimental design shown in Table 10).Results indicated that Test Level 3 (with minimal angle of repose error) represented the optimal region center point. Considering the error variation trend (showing "low-high" fluctuation from Levels 2 to 4), Levels 3 and 5 were selected as the low/high levels respectively for response surface model construction.The predetermined calibrated values were directly adopted for: salt-steel restitution coefficient, salt-steel/salt-salt static friction coefficients, and salt-salt Poisson's ratio.

Table 10 Steepest Climbing Test Plan and Results

| Number | X Salt-salt rolling friction | Y Salt-salt recovery coefficient | Z salt-steel rolling friction | angle of repose | relative error |
|---|---|---|---|---|---|
| 1 | 0.11 | 0.75 | 0.26 | 34.89 | 0.23 |
| 2 | 0.17 | 0.6 | 0.32 | 43.65 | 0.04 |
| 3 | 0.23 | 0.45 | 0.38 | 46.87 | 0.04 |
| 4 | 0.29 | 0.3 | 0.44 | 50.17 | 0.11 |
| 5 | 0.35 | 0.15 | 0.5 | 53.58 | 0.18 |

2.3.6 Box-Behnken Design and Regression Modeling

Building upon the parameter optimization region identified through steepest ascent experiments, a three-factor Box-Behnken design was implemented with quadratic regression orthogonal rotation: salt-salt restitution coefficient, salt-salt rolling friction coefficient, and salt-steel rolling friction coefficient as independent variables.Using the angle of repose ($R_1$) as the response variable, response surface analysis was conducted via Design-Expert 10.0.7 software (factor coding shown in Table 10, experimental design and results in Table 11, where X、Y、Z represent coded factor values), enabling precise identification of the optimal contact parameter combination.

Table 11: Coding of Simulation Test Factors

| Factor number | -1 | 0 | +1 |
|---|---|---|---|
| X | 0.17 | 0.23 | 0.29 |
| Y | 0.3 | 0.45 | 0.6 |
| Z | 0.32 | 0.38 | 0.44 |

Based on the data obtained from the simulation experiments, a multiple regression analysis was performed on the experimental results (Table 11) using DesignExpert 10.0.7 software, yielding the following quadratic regression equation for the angle of repose:

$$\theta = +45.57 + 1.73A - 0.67B + 0.43C - 0.66AC + 0.77BC + 1.23A^2 + 0.45B^2 \quad (28)$$

Table 11 Simulation Test Plan and Results

| Serial Number | X | Y | Z | angle of repose |
|---|---|---|---|---|
| 1 | -1 | -1 | 0 | 46.63 |
| 2 | +1 | -1 | 0 | 49.27 |
| 3 | -1 | +1 | 0 | 44.21 |
| 4 | +1 | +1 | 0 | 48.56 |
| 5 | -1 | -1 | -1 | 44.34 |
| 6 | +1 | -1 | -1 | 48.6 |
| 7 | -1 | -1 | +1 | 46.56 |
| 8 | +1 | -1 | +1 | 48.58 |
| 9 | 0 | -1 | -1 | 47.52 |
| 10 | 0 | +1 | -1 | 44.36 |
| 11 | 0 | -1 | +1 | 46.39 |
| 12 | 0 | +1 | +1 | 46.89 |
| 13 | 0 | -1 | 0 | 46.54 |
| 14 | 0 | -1 | 0 | 45.6 |
| 15 | 0 | -1 | 0 | 45.7 |
| 16 | 0 | -1 | 0 | 45.54 |
| 17 | 0 | -1 | 0 | 45.56 |

The ANOVA results for the Box-Behnken model (Table 12) revealed the following: The regression model was statistically significant at an extremely high level (P < 0.0001). The main effect terms (A, B, C), interaction term (BC), and quadratic term ($A^2$) were all highly significant (P < 0.01), while the interaction terms (AB, AC) reached significance (P < 0.05). Model validation metrics indicated: The lack-of-fit term was nonsignificant (P = 0.815), and the coefficient of variation (CV = 0.77%, <10%) confirmed good experimental repeatability. The coefficient of determination ($R^2$ = 0.98), adjusted $R^2$ ($R^2_{adj}$ = 0.95), and predicted $R^2$ ($R^2_{pre}$ = 0.90) all exceeded 0.9, demonstrating excellent model explanatory power and predictive capability. Additionally, the adequate precision value (17.96) indicated a high signal-to-noise ratio.

Table 12 Analysis of Variance

| Variance source | mean square | degrees of freedom | Sum of squared | F value | p value |
|---|---|---|---|---|---|
| Model | 38.72 | 9 | 4.3 | 33.74 | < 0.0001 |
| A | 22.01 | 1 | 22.01 | 172.6 | < 0.0001 |
| B | 4.19 | 1 | 4.19 | 32.86 | 0.0007 |
| C | 1.62 | 1 | 1.62 | 12.7 | 0.0092 |
| AB | 0.731 | 1 | 0.731 | 5.73 | 0.0479 |
| AC | 1.25 | 1 | 1.25 | 9.84 | 0.0165 |
| BC | 3.35 | 1 | 3.35 | 26.26 | 0.0014 |
| A² | 4.68 | 1 | 4.68 | 36.73 | 0.0005 |
| B² | 0.4441 | 1 | 0.4441 | 3.48 | 0.1043 |
| C² | 0.1323 | 1 | 0.1323 | 1.04 | 0.3424 |
| Residual | 0.8927 | 7 | 0.1275 | | |
| Lack of Fit | 0.1706 | 3 | 0.0569 | 0.3151 | 0.8151 |
| Pure Error | 0.7221 | 4 | 0.1805 | | |
| Cor Total | 39.62 | 16 | | | |
| $R^2=0.98$ | $R^2_{adj}=0.95$ | $R^2_{pre}=0.90$ | CV=0.77 | Adep precision=17.97 | |

2.3.7 Interaction Effect Analysis of Regression Model

The ANOVA of the Box-Behnken model indicated that the interaction effects between salt-salt rolling friction and salt-steel rolling friction (X×Z), as well as between salt-salt restitution coefficient and salt-steel rolling friction (Y×Z), were highly significant (P < 0.01). Meanwhile, the interaction between salt-salt rolling friction and restitution coefficient (X×Y) was significant (P < 0.05). Three-dimensional response surface analysis (Fig. 9) revealed the following:

Fig. 9a (X-Y interaction): When salt-steel rolling friction was fixed at the baseline level, the angle of repose increased with higher salt-salt restitution coefficients, with a more pronounced increase observed at elevated salt-salt rolling friction values.

Fig. 9b (Y-Z interaction): With the salt-salt restitution coefficient held constant at the baseline level, the angle of repose rose as salt-steel rolling friction increased, and this effect was significantly amplified at higher salt-steel rolling friction values.

These findings demonstrate a nonlinear regulatory mechanism governing granular pile formation under multi-parameter coupling effects.

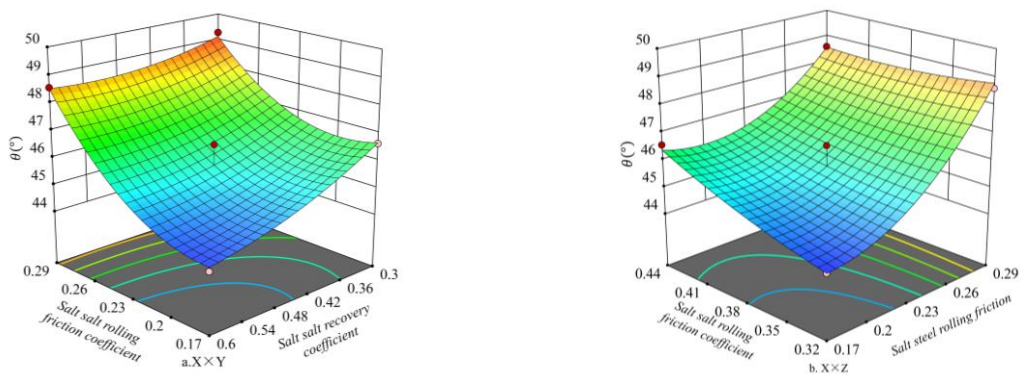

Figure 9 Response surface of the influence of interactive factors on simulation stacking angle

3 Experimental Validation

3.1 Optimal Parameter Determination

Using the experimentally measured angle of repose (45.27°, Table 13) as the target response value, the regression equation was solved through the optimization module of Design-Expert 10.0.7. Response surface analysis yielded multiple sets of optimal parameters, from which the following were ultimately selected: salt-salt rolling friction coefficient of 0.23, salt-salt restitution coefficient of 0.544, and salt-steel rolling friction coefficient of 0.368.

3.2 Angle of Repose Comparative Validation

The simulated angle of repose was compared with bench-scale experimental measurements. Verification was conducted using funnel tests and EDEM simulations, with results shown in Table 13. The funnel specifications and testing methodology matched those of the bench-scale experiments. Five replicate angle of repose tests yielded an average value of 45.27°. The selected optimal parameters were implemented in EDEM for angle of repose simulation, using previously calibrated parameters for salt-steel interactions. Five simulation replicates produced an average angle of 45.55°, showing a 0.6% relative error compared to experimental measurements. This close agreement demonstrates that the calibrated simulation results effectively match physical test outcomes, confirming that the calibration method successfully establishes correspondence between simulation models and actual salt particle physical characteristics. The calibrated contact parameters for livestock salt particles can be reliably used in discrete element method simulations, providing a foundation for subsequent numerical analyses.

Table 13 Results of Stacking Angle Test

| Serial Number | | | | | | mean value | Total mean |
|---|---|---|---|---|---|---|---|
| 1 | 46.6 | 46.61 | 44.29 | 45.6 | 46.16 | 45.852 | 45.852 | |
| 2 | 45.54 | 43.65 | 46.85 | 46.85 | 45.79 | 45.736 | 45.736 | |
| 3 | 44.12 | 45.09 | 45.08 | 45.08 | 45.21 | 44.916 | 44.916 | 45.55 |
| 4 | 45.61 | 46.41 | 45.68 | 45.68 | 46.59 | 45.994 | 45.994 | |
| 5 | 45.02 | 45.21 | 44.36 | 44.36 | 47.21 | 45.232 | 45.232 | |

4 Conclusions

1）In this study, a particle scaling method was employed to double the particle size of livestock salt, and the contact parameters of the scaled particles were calibrated using the discrete element method. Through Plackett-Burman experiments, three key parameters significantly affecting the angle of repose were identified: the salt-salt rolling friction coefficient, salt-salt restitution coefficient, and salt-steel rolling friction coefficient. The remaining parameters were calibrated through a combined approach of bench-scale experiments and simulations.

2）Based on Box-Behnken experimental design, a quadratic regression model was established between these three key parameters and the angle of repose. ANOVA results revealed that in addition to the primary effects of the three key parameters, the interaction between salt-salt rolling friction coefficient and restitution coefficient, as well as between salt-salt rolling friction coefficient and salt-steel rolling friction coefficient, exerted highly significant influences (P<0.01) on the angle of repose of scaled livestock salt particles.

3）Using the experimentally measured angle of repose as the optimization target, the regression equation was solved to obtain the optimal parameter combination: salt-salt rolling friction coefficient of 0.23, salt-salt restitution coefficient of 0.544, and salt-steel rolling friction coefficient of 0.368. Comparative experiments demonstrated no significant difference (P>0.05) between simulated and measured angles of repose, validating the feasibility of using response surface methodology for calibrating DEM simulation parameters.

4）Simulation tests using the optimized parameters yielded an angle of repose of 45.55°, showing only 0.6% error compared to the experimental measurement (45.27°) with no significant difference. These results confirm that the contact parameters obtained through particle scaling theory can be accurately applied in discrete element simulations of livestock salt particles.

Acknowledgments

   The project has been supported by the following projects

   The agricultural projects of the Xinjiang Production and Construction Corps (NYHXGG,2023AA403)

   The science and technology program projects of the Xinjiang Production and Construction Corps (2024AB047)

   The high-level talent research startup projects of Shihezi University (RCZK202310)

Data availability

   The data of this study are available from the corresponding author or first author upon reasonable request.

Declaration of Competing Interest

   The authors declare that they have no known competing financial interests or personal relationships that could have appeared to influence the work reported in this paper.